\documentclass[twocolumn,showpacs,prb,citeautoscript,amsmath,amssymb,floatfix]{revtex4}
\usepackage{dcolumn}
\usepackage{amsmath,amssymb}
\usepackage{bm}
\usepackage{epsfig}
\newcommand{\bd}{\bm}
\begin{document}
\title{Condensate density of  interacting 
bosons: a functional renormalization group approach} 
\author{Christopher Eichler$^1,$ Nils Hasselmann$^2$, and Peter Kopietz$^1$}
\vspace{2mm}  
\address{$^1$Institut f\"{u}r Theoretische Physik, Universit\"{a}t
    Frankfurt, Max-von-Laue-Strasse 1, 60438 Frankfurt, Germany\\
	$^2$International Center for Condensed Matter Physics, Universidade de Bras\'{\i}lia, 
Caixa Postal 04667, 70910-900 Bras\'{\i}lia, DF, Brazil}
%
\date{June 4, 2009}
\begin{abstract}
\noindent
We calculate the temperature dependent  condensate density
$\rho^0 ( T )$
of interacting bosons in three dimensions using the functional
renormalization group (FRG). From the numerical solution of
suitably truncated  FRG flow equations for the irreducible vertices
we obtain $\rho^0 ( T )$ for arbitrary temperatures. We carefully
extrapolate our numerical results to the critical point and determine
the order parameter exponent $\beta \approx 0.32$, in reasonable 
agreement with the expected value  $ 0.345$ associated with the XY-universality class.
We also calculate the condensate density in two dimensions at zero temperature
using a truncation of the FRG flow equations based on the derivative expansion
including  cubic  and quartic  terms in the expansion of the
effective potential in powers of the density.
As compared with  the widely used quadratic 
approximation for the effective potential, the coupling constants 
associated with the cubic and quartic terms 
increase the result for the condensate density by a few percent.
However, the cubic and quartic coupling constants  flow to rather large
values, which sheds some doubt on FRG
calculations based on a low order polynomial 
approximation for the effective potential.
\end{abstract}
\maketitle
\section{Introduction}
\label{sec:intro}

The condensed phase of interacting bosons is often studied using
Bogoliubov's celebrated mean-field 
approximation \cite{Bogoliubov47}.
However, if one tries to go beyond the Bogoliubov approximation
and  includes fluctuation corrections
perturbatively, some terms in the perturbation series
for the single-particle Green function
diverge. The upper critical dimension for these divergencies
is $D=3$ at zero temperature and $D=4$ at finite temperature.
In the critical dimension the divergencies are
logarithmic,~\cite{Nepomnyashchy75} whereas in lower dimensions
one encounters even stronger power law singularities.
Physically, these divergencies arise due to the coupling
of  transverse fluctuations to longitudinal ones in the condensed phase;
the gapless nature of the transverse fluctuations associated with the
Goldstone modes gives then rise to singularities in the
perturbation series of longitudinal correlation functions~\cite{Patashinskii73,Chakravarty91,
Giorgini99,Sachdev99b,Zwerger04,Kreisel07}, 
which have to be re-summed to all orders in perturbation theory to obtain meaningful results.
The controlled calculation of
physical properties of interacting
bosons  requires therefore non-perturbative methods.

Recently several  authors have studied interacting bosons by means of the
renormalization group~\cite{Castellani97,Dupuis07,Wetterich08,Sinner09,Dupuis09,Floerchinger09}, which is an efficient method to re-sum the perturbation series
and  remove the singularities encountered in finite order perturbation theory.
While most calculations so far have focused on
properties of  the superfluid ground-state \cite{Castellani97,Dupuis07,Wetterich08}
or on the single-particle spectral function at zero temperature \cite{Sinner09,Dupuis09},
Floerchinger and Wetterich \cite{Floerchinger09}  have 
used the non-perturbative functional renormalization group (FRG)
to calculate 
thermodynamic observables of the interacting Bose gas in three dimensions.
They used a  truncation of the formally exact FRG flow equation for
the generating functional $\Gamma$ of the irreducible vertices \cite{Wetterich93} 
based on the derivative expansion \cite{Berges02}, 
retaining field gradients and density fluctuations in the expansion of
the effective potential  up to second order.
In Sec.~\ref{sec:FRGvert} we shall present 
an alternative derivation of the resulting finite temperature flow equations based
on the vertex expansion~\cite{Morris94}.
In contrast to 
Ref.~[\onlinecite{Floerchinger09}], we shall then carefully analyze the
temperature dependence of the condensate density
$\rho^0 ( T )$ in the critical regime and extract the order parameter exponent
$\beta$ from the numerical solution of our truncated FRG flow equations.
Our result $\beta \approx 0.32$ 
is quite close to the expected value $ \beta \approx 0.345$ of
the three-dimensional XY-universality class \cite{Pelissetto02}.
We therefore conclude  that our simple
truncation of the vertex expansion yields 
quantitatively accurate results for the condensate density for all temperatures, including the
critical regime.

In order to estimate the effect of higher order many-body interactions, we shall
in Sec.~\ref{sec:derivative} go back
the derivative expansion approach \cite{Berges02,Floerchinger09}
and calculate the condensate density at vanishing temperature 
within a truncation to second order in the derivatives but to {\it{fourth order}}
in the expansion of the effective potential $U ( \rho )$ in powers of
density fluctuations $\rho - \rho^0$,
 \begin{equation}
 U ( \rho ) \approx U^{(0)} + 
 \sum_{ k=2}^{4} \frac{ U^{(k)}}{k!} ( \rho - \rho^0 )^k.
 \label{eq:Urhodef}
 \end{equation}
We find that the coupling constants $U^{(3)}$ and $U^{(4)}$ flow to
rather large values, which sheds some doubt on
the quantitative accuracy of calculations based on the  gradient expansion
with quadratic approximation for the effective potential.
Nevertheless, we find that the three- and four-body interactions
described by $U^{(3)}$ and $U^{(4)}$ have only a rather small effect on the
numerical value of the condensate density.

\section{FRG approach to the condensed Bose gas}
\label{sec:FRGvert}

\subsection{Vertex expansion of the FRG flow equation}

Starting point of our investigation is the
following Euclidean action describing  bosons with mass $m$
subject to a repulsive contact interaction $u_0$,
\begin{eqnarray}
S[\bar\psi,\psi]  & = & \int d^D r \int_0^{\beta} d \tau \Bigl[
\bar{\psi} ( \bd{r} , \tau ) (\partial_\tau-\frac{{\bd{\nabla}}^2}{2m}-\mu) \psi
 ( \bd{r} , \tau )
 \nonumber
 \\
& & \hspace{20mm}
+ \frac{u_0}{2} (\bar{\psi} ( \bd{r} , \tau )  {\psi} ( \bd{r} , \tau ))^2 \Bigr],
\label{eq:InAct}
\end{eqnarray}
where the chemical potential $\mu$ and the inverse temperature $\beta = 1/T$ are fixed.
The spatial integrals should be regularized by means of a short-distance 
cutoff $\Lambda^{-1}_0$, which is related to the finite extent of the interaction or,
for hard core bosons, to the size of the particles. 
The model~ (\ref{eq:InAct}) depends on three dimensionless parameters 
 \begin{subequations}
 \begin{eqnarray}
 \tilde \mu & = & \frac{2m \mu }{\Lambda_0^2},
 \\
\tilde T & = & 2 \pi \frac{2m T }{  \Lambda_0^2} ,
 \label{eq:tildeTdef}
 \\
\tilde u_0 & = & 2mu_0\Lambda^{D-2}_0,
 \label{eq:u0tildedef}
 \end{eqnarray}
 \end{subequations}
 where at this point we do not specify the dimensionality $D$ of the system,
and the factor of $2 \pi$
in the definition of the dimensionless temperature $\tilde{T}$ is introduced 
for later convenience.
We focus on the condensed phase, where the global $U(1)$-symmetry of the action
(\ref{eq:InAct}) is spontaneously broken and the field $\psi$ has a finite expectation value
$\phi^0 = \langle \psi ( \bd{r} , \tau ) \rangle$, which by translational invariance 
is independent of space $\bd{r}$ and imaginary time $\tau$.
Without loss of generality, we choose   $\phi^0$ to be real.

To derive formally exact FRG flow equations for the one-particle
irreducible vertices of our model, we add a cutoff dependent
regulator function $R_{\Lambda} (  \bd{k}  )$ to the inverse free propagator in the
Gaussian part of the action (\ref{eq:InAct}).
In momentum-frequency space the inverse free propagator is then
 \begin{equation}
 G^{-1}_{ 0 , \Lambda } ( K ) = 
 i \omega - \epsilon_{\bd{k}} + \mu - R_{\Lambda} (  \bd{k}  ),
 \end{equation}
where $i \omega$ is a bosonic Matsubara frequency,
$\epsilon_{\bd{k} } = \bd{k}^2 /2m$ is the free dispersion in momentum space,
and $K = ( \bd{k} , i \omega )$ is a collective label.
The regulator function  $R_{\Lambda} (  \bd{k}  )$
should satisfy 
 \begin{equation}
  R_{\Lambda} (  \bd{k}  ) \sim \left\{
 \begin{array}{cc} 0 & \mbox{for $\Lambda \rightarrow 0 $},
 \\
 \infty & \mbox{for $\Lambda \rightarrow \infty $},
 \end{array}
 \right.
 \end{equation}
so that the infrared cutoff $\Lambda$ suppresses long-wavelength fluctuations and
we recover our original model for $\Lambda \rightarrow 0$.
For convenience we use the Litim  regulator\cite{Litim01}
 \begin{equation}  
 R_\Lambda( \bd{k} )=\left( 1-\delta_{{\bm k},0} \right)
 Z_{\Lambda}^{-1}
\left( \epsilon_{\Lambda} - \epsilon_{\bd{k}} \right)
\Theta\left(\Lambda^2-\bd{k}^2\right),
 \label{eq:Litim} 
\end{equation}
where  the  dimensionless wave-function renormalization factor 
$Z_{\Lambda} $ is defined in Eq. (\ref{eq:approxsigma}) below.
The cutoff dependent  irreducible vertices 
$\Gamma^{ (n,m)}_{\Lambda} ( K_1^{\prime} , \ldots , K^{\prime}_n ; K_m, \ldots , K_1 )$
in the condensed phase are defined via the functional Taylor expansion
of the corresponding  generating functional \cite{footnotegenerating} in powers
of the fluctuations $\delta \phi_K = \phi_K - \delta_{ K,0} \phi^0$,
 \begin{eqnarray}
 \Gamma_{\Lambda} [ \bar \phi ,  \phi  ]
& = & \sum_{ n ,m =0}^{\infty} \frac{1}{ n! m!} \int_{ K_1^{\prime} }
 \cdots \int_{ K_n^{\prime} } \int_{ K_m }
 \cdots \int_{ K_1 } 
 \nonumber
 \\
 & \times & \delta_{ K_1^{\prime} + \ldots + K_n^{\prime} , K_m + \ldots + K_1 }
\nonumber
 \\
 & \times &
 \Gamma^{ (n,m)}_{\Lambda} ( K_1^{\prime} , \ldots , K^{\prime}_n ; K_m, \ldots , K_1 )
 \nonumber
 \\
 & \times & 
\delta \bar{\phi}_{ K_1^{\prime}} \ldots  \delta \bar{\phi}_{K_n^{\prime}}
\delta {\phi}_{ K_m} \ldots  \delta {\phi}_{K_1}.
 \label{eq:gendef}
 \end{eqnarray}
The derivative of the functional
$\Gamma_{\Lambda} [ \bar \phi ,  \phi  ]$
with respect to the infrared cutoff $\Lambda$ can be expressed
in closed form in terms of a deceptively simple
FRG flow equation \cite{Wetterich93,Berges02,Morris94}, which is equivalent to
an infinite hierarchy of 
integro-differential equations for the cutoff-dependent  vertices 
$\Gamma^{ (n,m)}_{\Lambda} ( K_1^{\prime} , 
\ldots , K^{\prime}_n ; K_m, \ldots , K_1 )$.
Following Refs.~[\onlinecite{Schuetz06,Sinner08,Kopietz09}], 
we fix the flowing order parameter 
$\phi^0_{\Lambda} = \langle \psi ( \bd{r} , \tau ) \rangle$
by demanding that the vertices 
$\Gamma^{(1,0)}_{\Lambda}$ and
$\Gamma^{(0,1)}_{\Lambda}$ with a single external leg 
should vanish identically for any value of the
cutoff $\Lambda$.

We are interested in the temperature dependent order parameter 
$\phi^0_{\ast} = \lim_{\Lambda \rightarrow 0} \phi^0_{\Lambda}$,
which determines the condensate density via $\rho^0_{\ast} = ( \phi^0_{\ast} )^2$.
The exact FRG flow equation 
for the flowing order parameter $\phi^0_{\Lambda}$
depends on the flowing normal and anomalous self-energies,\cite{Schuetz06}
 \begin{eqnarray}
      \Gamma^{(1,1)}_{\Lambda} ( K , K ) & = & 
 \Sigma^N_{\Lambda} ( K ),
 \label{eq:sigmaNdef}
 \\
\Gamma^{(0,2)}_{\Lambda} ( K , -K ) = \Gamma^{(2,0)}_{\Lambda} ( -K , K )  &  = &  
\Sigma^A_{\Lambda} ( K )  ,
\label{eq:sigmaAdef}
\end{eqnarray}
and on the four types of
vertices with three external legs, $\Gamma^{(3,0)}_{\Lambda}$, 
$\Gamma^{(2,1)}_{\Lambda}$, 
$\Gamma^{(1,2)}_{\Lambda}$ and
$\Gamma^{(0,3)}_{\Lambda}$.
To calculate the order parameter  we therefore need the flowing self-energies and
the flowing three-legged vertices, whose flow equations depend again on  
higher order vertices with four and more external legs.
To obtain a closed system of FRG flow equations, we shall use here the truncation
proposed in Ref.~[\onlinecite{Sinner09}], which amounts to 
the following 
parameterization of the non-zero vertices with three and four external legs,
 \begin{eqnarray}
 \Gamma^{(2,1)}_{\Lambda} ( K_1^{\prime} , K_2^{\prime} ; K_1 ) & = &
 \phi^0_{\Lambda} [ u_{\Lambda} ( K_1^{\prime} ) +   u_{\Lambda} ( K_2^{\prime} ) ],
 \label{eq:Gamma21trunc}
 \\
\Gamma^{(1,2)}_{\Lambda} ( K_1^{\prime} ;  K_2 , K_1 ) & = &
 \phi^0_{\Lambda} [ u_{\Lambda} ( K_1 ) +   u_{\Lambda} ( K_2 ) ],
 \hspace{7mm}
 \label{eq:Gamma12trunc}
\\
\Gamma^{(2,2)}_{\Lambda} ( K_1^{\prime} ,K_2^{\prime} ;  K_2 , K_1 ) & = & 
u_{\Lambda} ( K_1^{\prime} - K_1 )  + u_{\Lambda} ( K_2^{\prime} -K_1 ).
\nonumber
 \\
 & &
 \label{eq:Gamma22trunc} 
\end{eqnarray}
The system of flow equations for the
order-parameter and the self-energies is closed by demanding
that  the momentum- and frequency dependent function $u_{\Lambda} ( K )$
is related to the flowing self-energies via 
\begin{subequations}
  \begin{eqnarray}
    \Sigma_\Lambda^N(K) & = & \sigma_\Lambda(K)
+ \rho_\Lambda^0 [u_\Lambda(0)+u_\Lambda(K)]  , 
    \label{eq:approxsigman}\\
    \Sigma_\Lambda^A(K) & = & \rho_\Lambda^0 u_\Lambda(K) ,
    \label{eq:approxsigmaa}
  \end{eqnarray}
\end{subequations}
where $\rho_{\Lambda}^0 = (\phi_{\Lambda}^0 )^2$ is the flowing condensate density and
$\sigma_{\Lambda} ( K )$ is another $K$-dependent function
satisfying $\sigma_{\Lambda} ( 0 ) = \mu - \rho_{\Lambda}^0 u_{\Lambda} ( 0 )$.
Eqs.~(\ref{eq:Gamma21trunc}--\ref{eq:Gamma22trunc}) relate the 
vertices with three and four external legs to the normal and anomalous components
of the irreducible self-energy, and thus close the FRG flow equations for the 
order parameter and the self-energies.
Our parameterization of the self-energies given in 
Eqs.~(\ref{eq:approxsigman}, \ref{eq:approxsigmaa})
is motivated by the pioneering insights gained by
Nepomnyashchy and Nepomnyashchy~\cite{Nepomnyashchy75}, implying that
only the contribution to the self-energy contained in the function
 $u_{\Lambda} ( K )$ exhibits a  non-analytic $K$-dependence for
 $\Lambda \rightarrow 0$,
while the contribution $\sigma_{\Lambda} ( K )$ remains analytic.
As noted in Ref.~[\onlinecite{Sinner09}], the above truncation
satisfies the Hugenholtz-Pines relation~\cite{Hugenholtz59},
 \begin{equation}
 \Sigma^N_{\Lambda} (0)  -  \Sigma^A_{\Lambda}  ( 0) = \mu,
 \end{equation}
as well as the Nepomnyashchy identity~\cite{Nepomnyashchy75}, 
\begin{equation}
 \lim_{\Lambda \rightarrow 0 }   \Sigma^A_{\Lambda} (0)  =0,
 \end{equation}
which holds at finite temperature for $D \leq 4$ and at zero temperature for
$D \leq 3$, since in these cases
\begin{equation}
\lim_{\Lambda \rightarrow 0} u_{\Lambda}(0)= 0.
\end{equation}
The truncation (\ref{eq:Gamma21trunc}--\ref{eq:Gamma22trunc}) amounts to the 
following approximation for the generating functional $\Gamma_{\Lambda} [ \bar{\phi},
 \phi ]$ defined in Eq.~(\ref{eq:gendef}),
\begin{equation}
\Gamma_\Lambda  [ \bar{\phi}, \phi ]
= \Gamma^{(0)}_{\Lambda} + \int_K \bar{\phi}_K \sigma_\Lambda(K)\phi_K 
+ \frac{1}{2} \int_K \rho_K u_\Lambda(K) \rho_{-K},
\label{eq:defgamma}
\end{equation} 
where $\rho_K=\int_Q \bar{\phi}_{Q}\phi_{Q+K}$ are the Fourier components of the
density~\cite{footnotemistake},
and $\Gamma^{(0)}_{\Lambda}$ is an interaction correction to the
grand canonical potential in units of the temperature.
As shown in Ref.~[\onlinecite{Sinner09}], 
 for $\Lambda \rightarrow 0$  the function
$u_{\Lambda} ( K )$ develops 
a non-analytic dependence on $K$, so that
the corresponding effective potential in real space and imaginary time
is non-local.

To further simplify the FRG flow equations, we shall 
replace on the right-hand sides of the flow equations,
 \begin{equation}
u_{\Lambda} ( K ) \rightarrow u_{\Lambda} ( 0 ) = u_{\Lambda}.
 \end{equation}
Within this truncation, the FRG flow equation for the  condensate density
$\rho^0_{\Lambda} = (\phi^0_{\Lambda} )^2$ reduces to
 \begin{equation}
 \partial_{\Lambda} \rho_{\Lambda}^0 = \int_Q [ 2 \dot{G}^N_{\Lambda} ( Q ) 
 +  \dot{G}^A_{\Lambda} ( Q )  ],
 \label{eq:rhoflow}
 \end{equation}
while the normal and anomalous components of the self-energy satisfy~\cite{Sinner09}
\begin{widetext}
    \begin{eqnarray}
      \partial_\Lambda\Sigma^{N}_\Lambda(K)& = & 2u^{}_\Lambda\int_{Q}\left\{ 
        \dot{G}^{N}_\Lambda(Q)+\dot{G}_\Lambda^{A}({Q})\right\} 
      \nonumber
      -  4u_\Lambda^2 \rho^0_\Lambda\int_{Q}\Big\{ \dot{G}_\Lambda^{N}({Q}) 
      \big[G_\Lambda^{N}(Q+K)+G_\Lambda^{N}(Q-K)+G_\Lambda^{N}(-Q+K)       
      \\
      & & + 2 G^{A}_\Lambda(Q-K) \big]
      +2\dot G_\Lambda^{A}(Q) 
      \big[G_\Lambda^{A}(Q+K) + G_\Lambda^{N}(Q+K) \big] \Big\} \, ,       
      \label{eq:NormVertFlow} 
      \\
      \partial_\Lambda\Sigma^{A}_\Lambda(K)& = & 2u_\Lambda\int_{Q}
      \dot{G}^{N}_\Lambda(Q) 
      -4 u_\Lambda^2 \rho^0_\Lambda \int_Q \Big\{ \dot{G}^{N}_\Lambda(Q) \big[G_\Lambda^{N}(Q+K)
      +G_\Lambda^{N}(Q-K)+G_\Lambda^{A}(Q+K)
      +G_\Lambda^{A}(Q-K) \big] \nonumber 
      \\
      && +  \dot{G}^{A}_\Lambda(Q) \big[G_\Lambda^{N}(Q+K)+G_\Lambda^{N}(Q-K)
      +3G_\Lambda^{A}(Q+K) \big] \Big\} \, .
      \label{eq:AnomVertFlow}
    \end{eqnarray}
\end{widetext}
Here the single-scale propagators $\dot{G}_\Lambda^{N}(K)$ and
$\dot{G}_\Lambda^{A}(K)$ are defined via the matrix equation
 \begin{equation}
 \left(
    \begin{array}{cc}
      \dot{G}_\Lambda^N(K) & \dot{G}_\Lambda^A(K) \\
      \dot{G}_\Lambda^A(K)^* & \dot{G}_\Lambda^N(-K)
    \end{array}
  \right) 
=-{\bm G}_\Lambda(K) 
[ \partial_\Lambda{\bm G}_{0,\Lambda}^{-1}(K)] {\bm G}_\Lambda(K),
 \end{equation}
where  $  {\bm G}_{\Lambda}^{-1}(K)  
=   {\bm G}_{0,\Lambda}^{-1}(K)-{\bm \Sigma_\Lambda}(K)$,
and
\begin{eqnarray}
  {\bm G}_{0,\Lambda}^{}(K)&=&\left(
    \begin{array}{cc}
      G_{0,\Lambda}^{}(K) & 0\\
      0 & G_{0,\Lambda}^{}(-K)
    \end{array}
  \right) ,  \\
  {\bm \Sigma}_\Lambda(K)&=&\left(
    \begin{array}{cc}
      \Sigma_\Lambda^N(K) & \Sigma_\Lambda^A(K) \\
      \Sigma_\Lambda^A(K)^* & \Sigma_\Lambda^N(-K)
    \end{array}
  \right) .
\end{eqnarray}

\subsection{Low-energy truncation and results}
 \label{sec:lowenergy}
Following Ref.~[\onlinecite{Sinner09}] we expand the analytic part 
 $\sigma_{\Lambda} ( K )$ of the self-energy in powers of momenta and frequencies up
to quadratic order,
  \begin{eqnarray}
    \sigma_\Lambda(K) &\approx&  
\mu  ( 1 - X_{\Lambda} )  + i \omega (1-Y_\Lambda) 
 \nonumber \\
 &+& \epsilon_{ \bd{k}}(Z_\Lambda^{-1}-1)
      - (i \omega)^2 V_\Lambda  \, , 
    \label{eq:approxsigma} 
  \end{eqnarray}
where $X_{\Lambda} = \rho_\Lambda^0 u_{\Lambda} / \mu$.
Note that at the initial scale 
$X_{\Lambda_0} = Y_{\Lambda_0} = Z_{\Lambda_0} =1$
and $V_{\Lambda_0} =0$, so that
  $\sigma_{\Lambda_0}(K)$ vanishes.
While at zero temperature it is essential to retain 
the couplings $Y_{\Lambda}$ and $V_{\Lambda}$
associated with the frequency dependence of the self-energy, 
these couplings are irrelevant at the critical fixed point associated with
Bose-Einstein condensation, which is a classical phase transition.
Since in this section we are interested in the finite temperature behavior of the
condensate density, for our purpose it is sufficient to 
retain only the couplings $X_{\Lambda}$ and
$Z_{\Lambda}$ in Eq.~(\ref{eq:approxsigma}),
 \begin{equation}
    \sigma_\Lambda(K) \approx \sigma_{\Lambda} ( \bd{k} , i \omega =0) \approx \mu - r_{\Lambda} +  
 \epsilon_{ \bd{k}}  (Z_\Lambda^{-1}-1),
 \end{equation}
where we have introduced the notation
 \begin{equation}
 r_{\Lambda} = \rho_{\Lambda}^0 u_{\Lambda}.
 \end{equation}
With this truncation, the normal and anomalous propagators are simply
\begin{eqnarray}
G^{N}_{\Lambda}(K) &=& \frac{ -i\omega  - Z^{-1}_{\Lambda} 
\epsilon_{\bd{k}} -r_{\Lambda}  + R_{\Lambda} ( \bd{k} )  }{ \omega^2 -
 r_{\Lambda}^2  + [ Z^{-1}_{\Lambda} \epsilon_{\bd{k}} + r_{\Lambda}   -  
R_{\Lambda} ( \bd{k} )   ]^2} ,
 \label{eq:GNapprox}
 \\
G^{A}_{\Lambda}(K) &=& 
\frac{ r_{\Lambda}  }{ \omega^2 -
 r_{\Lambda}^2  + [ Z^{-1}_{\Lambda} \epsilon_{\bd{k}} + r_{\Lambda}   -  
R_{\Lambda} ( \bd{k} )   ]^2} ,
 \hspace{7mm}
\label{eq:GAapprox} 
\end{eqnarray}
while the corresponding  single-scale propagators are
  \begin{eqnarray}
\dot{G}^{N}_{\Lambda} (K) &=& - [ \partial_{\Lambda}  R_{\Lambda} ( \bd{k} ) ]
 \nonumber
 \\
 & & \hspace{-15mm} \times
\frac{ r_{\Lambda}^2 +  [  -i\omega  - Z^{-1}_{\Lambda} 
\epsilon_{\bd{k}} -r_{\Lambda}  + R_{\Lambda} ( \bd{k} ) ]^2 
  }{ \bigl[ \omega^2 -
 r_{\Lambda}^2  + [ Z^{-1}_{\Lambda} \epsilon_{\bd{k}} + r_{\Lambda}   -  
R_{\Lambda} ( \bd{k} )   ]^2 \bigr]^2 } ,
 \label{eq:dotGNapprox}
 \\
\dot{G}^{A}_{\Lambda} (K) &=& - [ \partial_{\Lambda}  R_{\Lambda} ( \bd{k} ) ]
 \nonumber
 \\
 & & \hspace{-15mm} \times
\frac{ 2 r_{\Lambda} [ R_{\Lambda} ( \bd{k} )  - Z^{-1}_{\Lambda}  
\epsilon_{\bd{k}} -r_\Lambda  ]    
}{ \bigl[ \omega^2 -
 r_{\Lambda}^2  + [ Z^{-1}_{\Lambda} \epsilon_{\bd{k}} + r_{\Lambda}   -  
R_{\Lambda} ( \bd{k} )   ]^2 \bigr]^2 } .
\label{eq:dotGAapprox} 
\end{eqnarray}
Substituting these expressions into Eqs.~(\ref{eq:rhoflow}--\ref{eq:AnomVertFlow})
and using
the Litim regulator (\ref{eq:Litim}), the momentum integrations and
frequency summations appearing in the FRG flow equations 
for  the three independent couplings $\rho^0_{\Lambda}$, $u_{\Lambda}$ and
$Z_{\Lambda}$ at finite temperature can all be performed analytically.
In $D$ dimensions the final result can be written in the following form,
\begin{eqnarray}
\Lambda \partial_{\Lambda} \tilde{\rho}^0_{\Lambda}
  &=&  \frac{ 2 K_D}{\pi D}  \left( \frac{\Lambda}{\Lambda_0} \right)^{D+2}  
 \left( 1 - \frac{\eta_{\Lambda}}{D+2} \right) Z_{\Lambda}^{-1}
 \nonumber
 \\
 &  \times &
 \left[
\tilde{\beta}^3  S_{0,2} ( \tilde{\beta} \tilde{E}_{\Lambda} )  
  P_{\Lambda}^{(2)}  - 
\tilde{\beta}  S_{2,2} ( \tilde{\beta} \tilde{E}_{\Lambda} )
 \right],
 \label{eq:rhoflowT}
\\
\Lambda \partial_{\Lambda} \tilde{u}_{\Lambda}
  &=&  - \frac{ 8 K_D}{\pi D}  \left( \frac{\Lambda}{\Lambda_0} \right)^{D+2}  
 \left( 1 - \frac{\eta_{\Lambda}}{D+2} \right) \tilde{u}_{\Lambda}^2 Z_{\Lambda}^{-1}
 \nonumber
 \\
 &   & \hspace{-10mm} \times
 \left[
\tilde{\beta}^5  S_{0,3} ( \tilde{\beta} \tilde{E}_{\Lambda} )  
P_{\Lambda}^{(3)} - 
\tilde{\beta}^3  S_{2,3} ( \tilde{\beta} \tilde{E}_{\Lambda} ) P_{\Lambda}^{(1)}
 \right],
 \label{eq:uflowT}
 \end{eqnarray}
\begin{equation}
\Lambda \partial_\Lambda Z_\Lambda =  \eta_{\Lambda} Z_{\Lambda},
\label{eq:ZflowT}
\end{equation}
where the flowing anomalous dimension is 
\begin{equation}
 \eta_{\Lambda} = \frac{ 4 K_D}{\pi D}  \left( \frac{\Lambda}{\Lambda_0}
 \right)^{D+2}  
\tilde{\rho}^0_{\Lambda} \tilde{u}_{\Lambda}^2 Z_{\Lambda}^{-1} 
\tilde{\beta}^3  S_{0,2} ( \tilde{\beta} \tilde{E}_{\Lambda} )  .
 \label{eq:etadef}
 \end{equation}
Here
$K_D=2^{1-D}\pi^{-D/2}/\Gamma[D/2]$ is the surface of the $D$-dimensional unit sphere
divided by $(2 \pi )^D$, the dimensionless inverse temperature
 $\tilde{\beta} = 1/ \tilde{T}$ is 
defined via Eq.~(\ref{eq:tildeTdef}), and
we have introduced dimensionless quantities
 \begin{subequations}
 \begin{eqnarray}
 \tilde{\rho}_{\Lambda}^0 & = & \rho_{\Lambda}^0 \Lambda_0^{-D},
 \\
 \tilde{u}_{\Lambda} & =  & 2 m u_{\Lambda} \Lambda_0^{D-2},
 \\
 \tilde{r}_{\Lambda} & = & \tilde{\rho}_{\Lambda}^0 \tilde{u}_{\Lambda} =
 2 m  \rho_{\Lambda}^0 u_{\Lambda} /  \Lambda_0^2 ,
 \\
  \tilde{E}_{\Lambda} & = & \sqrt{ \tilde{\epsilon}_{\Lambda} ( 
\tilde{\epsilon}_{\Lambda} + \tilde{r}_{\Lambda} ) } ,
 \\
 \tilde{\epsilon}_{\Lambda} & = & Z_{\Lambda}^{-1} \Lambda^2 / \Lambda_0^2,
 \end{eqnarray}
 \end{subequations}
as well as dimensionless coefficients,
 \begin{subequations}
\begin{eqnarray}
P_\Lambda^{(1)} & = &   \frac{7}{4}\tilde{r}_{\Lambda} +
\frac{11}{4}\tilde{\epsilon}_{\Lambda},
 \\
P_\Lambda^{(2)} & = &  \tilde{r}_\Lambda^2 + 
\tilde{r}_\Lambda \tilde{\epsilon}_{\Lambda} + \tilde{\epsilon}_{\Lambda}^2,
\\
P_\Lambda^{(3)} & = &  \tilde{r}_{\Lambda}^3 + \frac{3}{2} 
\tilde{r}_{\Lambda}^2 \tilde{\epsilon}_{\Lambda} + 
\frac{3}{4}\tilde{r}_{\Lambda} \tilde{\epsilon}_{\Lambda}^2 + 
\frac{5}{4}\tilde{\epsilon}_{\Lambda}^3.
\label{eq:Konstanten}
\end{eqnarray}
\end{subequations}
Finally, the dimensionless functions $S_{ k, l} ( x )$ are defined in terms of the
bosonic Matsubara sums,
\begin{eqnarray}
S_{ k, l} ( x ) = \sum_{n=-\infty}^{\infty} \frac{n^k}{(n^2 +x^2 )^l},
\end{eqnarray}
which can be expressed in terms of the Bose function and its derivatives.

\begin{figure}[h]
\centering
\includegraphics[width=8cm]{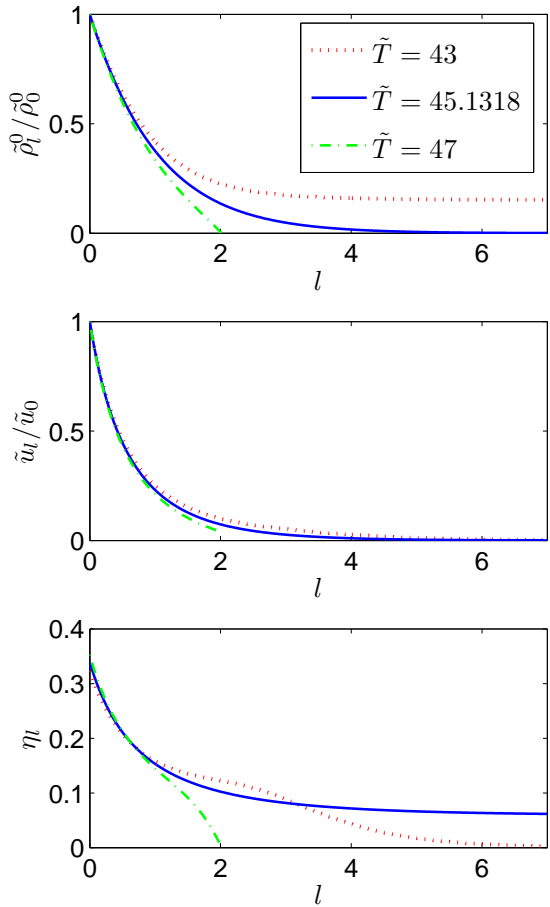}
\caption{(Color online) Typical FRG flow of the condensate density, the anomalous dimension 
and the dimensionless interaction $\tilde{u}_l$
for $D=3$ and $\tilde{\mu}  = \tilde{u}_0 = 1$.
Solid line: critical
temperature ($\tilde{T}_c = 45.1318$ for our choice of parameters);
dashed line: $\tilde{T} = 47 > \tilde{T}_c$; dotted line: $\tilde{T} = 43 < \tilde{T}_c$.
 } 
\label{fig:rhoetafluss}
\end{figure}

\begin{figure}[h]
\centering
\includegraphics[width=8cm]{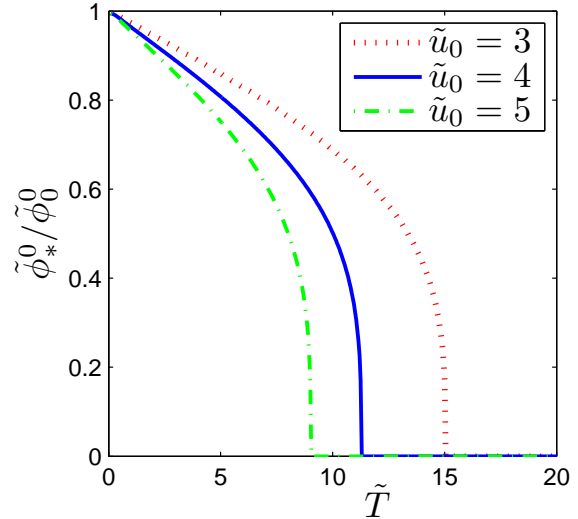}
\caption{ (Color online)  
Temperature dependence of the
order parameter  $\tilde{\phi}^0_{\ast} = \sqrt{ \tilde{\rho}^0_{\ast} }$
in three dimensions for $\tilde{\mu}=1$ and three different values of the
dimensionless bare interaction $\tilde{u}_0$ defined in
Eq.~(\ref{eq:u0tildedef}).
} 
\label{fig:orderpara}
\end{figure}

\begin{figure}[h]
\centering
\includegraphics[width=8cm]{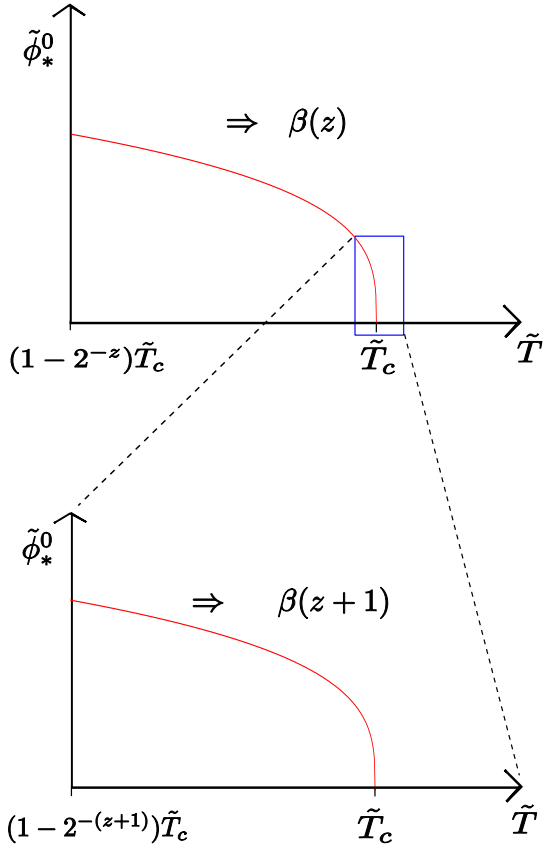}
\caption{(Color online)  
Iterative procedure to extract the
order parameter exponent $\beta$ from the
FRG results. We fit our numerical FRG results in a series of
intervals $I_z$
given by Eq.~(\ref{eq:interval}) to a power law (\ref{eq:powerlaw}) and thus
determine the critical exponent $\beta(z)$ for a given value of the
zoom factor $z$.
The $z$-dependence of $\beta(z)$ is then extrapolated 
for $z \rightarrow \infty$ as shown in Fig.~\ref{fig:betapol}.
 } 
\label{fig:betafit}
\end{figure}

The system of first order differential equations given by 
Eqs.~(\ref{eq:rhoflowT}--\ref{eq:ZflowT}) can easily be solved numerically.
It is convenient to consider quantities as functions of the logarithmic
flow parameter $l = - \ln ( \Lambda / \Lambda_0 )$, renaming
$ \tilde{\rho}^0_{\Lambda_0 e^{-l} } \rightarrow \tilde{\rho}_l^0$, and analogously for the 
other couplings.
In Fig.~\ref{fig:rhoetafluss} we show the typical RG flow
of the couplings $\tilde{\rho}_l^0$, $\tilde{u}_l$ and the flowing anomalous
dimension $\eta_l$ for three different temperatures (above, below and at the critical
temperature).
As expected, only at the critical temperature 
the flowing anomalous dimension has a finite limit (which can be identified with
the critical exponent $\eta$), while for $T < T_c$ the  
condensate density approaches a non-zero value
$\tilde{\rho}^0_{\ast}$. The temperature dependence
of the corresponding order parameter $\tilde{\phi}^0_{\ast} = \sqrt{ \tilde{\rho}^0_{\ast} }$
is shown in Fig.~\ref{fig:orderpara} for three different values of the interaction
strength.
In order to extract the order parameter exponent $\beta$ from
our numerical results shown in Fig.~\ref{fig:orderpara}, one should carefully
fit the curves in a sufficiently small temperature interval 
below the critical temperature to a power law,
 \begin{equation}
 \tilde{\phi}^0_{\ast} \propto ( \tilde{T}_c - \tilde{T} )^{\beta}.
 \label{eq:powerlaw} 
\end{equation} 
Due to the  lack of a priori knowledge about the proper 
interval for the power-law fit, and due to the finite accuracy 
of the numerical data, it is non-trivial to extract the
critical exponent $\beta$ from the numerical solutions of the
FRG flow equations shown in Fig.~\ref{fig:rhoetafluss}.
In fact, in a recent FRG calculation \cite{Floerchinger09}
of the temperature dependent condensate density
the critical exponent $\beta$
was not determined, apparently due to a lack of numerical
accuracy. Here we present an extrapolation  procedure
which allows us to obtain the critical exponent 
$\beta$ with high accuracy.
The crucial point is that one should use a series of  increasingly narrow
intervals close to the critical point for the fitting procedure.
Specifically, we use intervals of the form
 \begin{equation}
 I_z = [  (1 - 2^{-z} ) \tilde{T}_c  ,   \tilde{T}_c],
 \label{eq:interval}
 \end{equation}
which are parameterized in terms of the zoom factor $z$. 
For increasing values of $z$ we  fit 
the data in the corresponding temperature interval $I_z$ 
to the power law (\ref{eq:powerlaw})
and extract $\beta ( z )$.
This procedure is illustrated in Fig.~\ref{fig:betafit}.
We then plot our results for $\beta(z)$ as a function of $z$ and 
extrapolate for $z \rightarrow \infty$. As shown in Fig.~\ref{fig:betafit},
the dependence of  $\beta(z)$ on the zoom factor $z$ can be described by
 \begin{equation}
 \beta(z) \approx \beta [ 1 - e^{ - \alpha ( z + \gamma )} ],
 \label{eq:betazoom}
\end{equation}
with some non-universal numbers $\alpha$ and $\gamma$.
\begin{figure}[h]
\centering
\includegraphics[width=8cm]{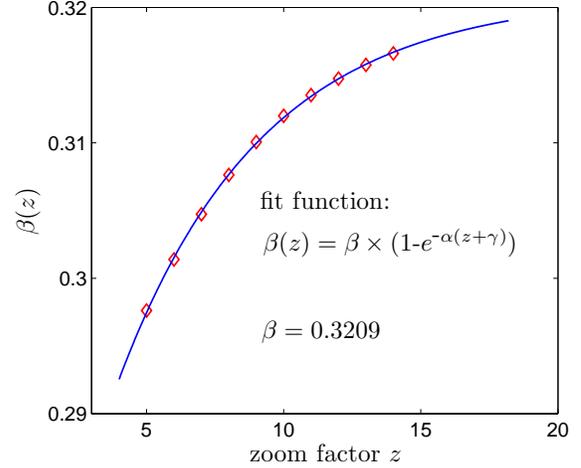}
\caption{(Color online)  
The dependence of $\beta(z)$ on the zoom factor $z$
can be described by the  indicated fit function
[see also Eq.~(\ref{eq:betazoom})] with
$\alpha = 0.1945$ and $\gamma = 8.545$.
The  data are based on the numerical solution
of the FRG flow equations for $\tilde{u}_0 = \tilde{\mu} =1$.
}
\label{fig:betapol}
\end{figure}
The critical exponent $\beta$ can then be identified with
$\beta = \lim_{z \rightarrow \infty} \beta(z)$.
Given the simplicity of our truncation
our final result $\beta \approx 0.32$
is in reasonable agreement
with the accepted value $\beta \approx 0.345$ of the three-dimensional
XY-universality class.~\cite{Pelissetto02}
We have checked that our extrapolated result for $\beta$ 
is independent of the various non-universal parameters
of our model, such as $\Lambda_0$ or the value of $\tilde{u}_0$.
We therefore believe that, in spite of its simplicity, our truncation of the
vertex expansion
yields accurate results for
the condensate density for all temperatures.

\section{Derivative expansion with quartic effective potential}
\label{sec:derivative}

The truncation of the vertex expansion used in the previous section
is equivalent to approximating the generating functional 
$\Gamma_{\Lambda} [ \bar{\phi} , \phi ]$ of the irreducible vertices   
by Eq.~(\ref{eq:defgamma}). 
A weak point of this truncation is that it neglects many-body interactions
involving more than two particles, encoded in the irreducible vertices with more
than four external legs.
Within the framework of the vertex expansion it is rather difficult to
take these vertices into account. In this section we shall therefore use the derivative 
expansion\cite{Berges02} with cubic and quartic terms in the effective potential to
estimate the effect of higher order many-body interactions.
For simplicity we consider in this section
only the quantum renormalization of the condensate density
at vanishing temperature.

If we approximate $u_{\Lambda} ( K ) \approx 
u_{\Lambda} ( 0 ) \equiv u_{\Lambda}$ 
 in Eq.~(\ref{eq:defgamma})
and use the low-energy expansion
(\ref{eq:approxsigma}) for the analytic part $\sigma ( K )$ of the self-energy,
then our truncated vertex expansion
of Sec.~\ref{sec:lowenergy}
 amounts to the following approximation for the
generating functional of the irreducible vertices,
 \begin{eqnarray}
& & \Gamma_{\Lambda} [ \bar{\phi} , \phi ] 
 - \mu \int d^D r \int_0^{\beta} d \tau \rho  ( \bd{r} , \tau )=
 \nonumber
 \\
 &  & 
\int d^D r \int_0^{\beta} d \tau   \Bigl[     {U}_{\Lambda} ( \rho ( \bd{r} , \tau ) )  
+  (1 - Y_{\Lambda}) \bar{\phi} ( \bd{r} , \tau ) \partial_{\tau} \phi ( \bd{r} , \tau )
\nonumber
 \\
& &    + 
 ( Z_{\Lambda}^{-1} -1 ) \frac{ | \bd{\nabla} \phi ( \bd{r} , \tau ) |^2}{ 2m}
 + V_{\Lambda}  | \partial_{\tau} \phi ( \bd{r} , \tau )  |^2 \Bigr],
 \label{eq:Gammagrad}
\end{eqnarray}
where   $ \rho  ( \bd{r} , \tau ) = | \phi ( \bd{r} , \tau ) |^2$,   and we have used the identity 
 \begin{equation}
 \mu ( 1 - X_{\Lambda} ) \rho + \frac{u_{\Lambda}}{2} \rho^2  
= \mu \rho  + \frac{u_{\Lambda}}{2} ( \rho - \rho^0_{\Lambda} )^2
  - \frac{u_{\Lambda}}{2} ( \rho^0_{\Lambda})^2,
 \end{equation}
to write the local effective potential in the form
 \begin{eqnarray}
 {U}_{\Lambda} ( \rho ) & = & U^{(0)}_{\Lambda} + 
\frac{u_{\Lambda}}{2} ( \rho - \rho^0_{\Lambda} )^2.
 \end{eqnarray}
Here
 \begin{equation}
  U^{(0)}_{\Lambda} = \frac{ \Gamma_{\Lambda}^{(0)}}{\beta {\cal{V}} } 
 - \frac{u_{\Lambda}}{2} ( \rho^0_{\Lambda})^2,
 \end{equation}
where ${\cal{V}}$ is the volume of the system.
Recall that throughout this work we normalize 
 $\Gamma_{\Lambda} [ \bar{\phi} , \phi ]$ such that 
it reduces to the two-body part of the bare action
at the initial RG scale~\cite{footnotegenerating}, 
where $X_{\Lambda_0} = Y_{\Lambda_0} = Z_{\Lambda_0} =1$ and
$V_{\Lambda_0} = \Gamma_{\Lambda_0}^{(0)} = 0$
(the chemical potential term is included in the non-interacting Green 
function).

\begin{figure}[h]
\centering
\includegraphics[width=8cm]{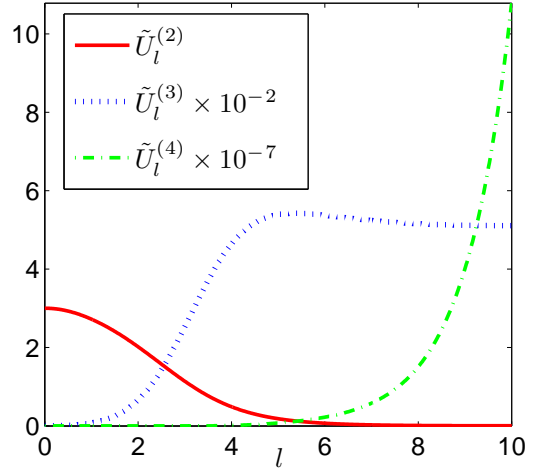}
\caption{(Color online)  
RG flow of the dimensionless couplings 
$\tilde{U}_{\Lambda}^{(2)}, 
\tilde{U}_{\Lambda}^{(3)}$ and
$\tilde{U}_{\Lambda}^{(4)}$
defined in Eq.~(\ref{eq:tildeU234def}) for $\tilde{u}_0=3$, $\tilde{\mu}=1$, $D=2$ and $T=0$.
} 
\label{fig:U234flow}
\end{figure}

To investigate the effect of higher order many-body interactions,
we now generalize the above ansatz by
replacing the effective potential by a fourth order polynomial in the density
 \begin{eqnarray}
 U_{\Lambda} ( \rho ) & = & U_\Lambda^{(0)} + 
 \frac{U_{\Lambda}^{(2)}}{2!} ( \rho - \rho^0_{\Lambda} )^2
 \nonumber
 \\
 &+ &  
\frac{U_{\Lambda}^{(3)}}{3!} ( \rho - \rho^0_{\Lambda} )^3
+\frac{U_{\Lambda}^{(4)}}{4!} ( \rho - \rho^0_{\Lambda} )^4.
 \end{eqnarray}
In fact, for the truncated generating functional  $\Gamma_{\Lambda} [ \bar{\phi} , \phi ] $
of the form (\ref{eq:Gammagrad}) with an arbitrary local effective potential
$U_{\Lambda} ( \rho )$ the exact FRG flow equation~\cite{Wetterich93,Berges02}
implies the following
partial differential equation for the effective potential \cite{Floerchinger09},
 \begin{eqnarray}
 \partial_{\Lambda} U_{\Lambda} ( \rho ) & = & 
\frac{K_D}{D} \left( 1 - \frac{\eta_{\Lambda}}{D+2} \right) \frac{\Lambda^{D+1}}{2m Z_{\Lambda}}
 \nonumber
 \\
 &  & \hspace{-20mm} \times \int \limits_{ - \infty}^{\infty} \frac{ d \omega}{2 \pi}
 \frac{ \rho U^{\prime \prime} ( \rho ) + U^{\prime} ( \rho ) + V_{\Lambda} \omega^2 }{
[ \rho U^{\prime \prime} ( \rho ) + U^{\prime} ( \rho ) + V_{\Lambda} \omega^2]^2 -
 [ \rho U^{\prime \prime} ( \rho ) ]^2 + Y^2_{\Lambda} \omega^2 }.
 \nonumber
 \\
 & &
 \label{eq:Udif}
 \end{eqnarray}
At finite temperature, the frequency integral should be replaced by a bosonic Matsubara sum,
 $  \int \frac{ d \omega}{2 \pi} \rightarrow
T \sum_{ \omega}$.
To obtain an approximate solution of the  partial differential equation (\ref{eq:Udif}),
we expand $U_{\Lambda} ( \rho )$ in powers of $\rho - \rho^0_{\Lambda}$. 
The flowing condensate density $\rho^0_{\Lambda}$ is then determined by
 \begin{equation}
  \left. \frac{ \partial U_{\Lambda} ( \rho )}{ \partial \rho } 
\right|_{ \rho^0_{\Lambda}} =0,
 \end{equation}
and the expansion coefficients are
\begin{equation}
U_{\Lambda}^{(0)} = U_{\Lambda} ( \rho^0_{\Lambda} ) \; \; \;  , \; \; \;  
  U_{\Lambda}^{(k)} = \left. \frac{ \partial^{k} U_{\Lambda} ( \rho )}{ \partial \rho^k} 
  \right|_{ \rho^0_{\Lambda}}.
 \end{equation}
Taking derivatives of Eq.~(\ref{eq:Udif}), we obtain the flow equations for the
condensate density and expansion coefficients,
\begin{eqnarray}
 \partial_{\Lambda} \rho^0_{\Lambda} &=& - \frac{1}{U_{\Lambda}^{(2)}} \frac{\partial}{\partial \rho} \big(\partial_{\Lambda} U_{\Lambda}(\rho)\big)\big|_{\rho^0_{\Lambda}},
\\ 
\partial_{\Lambda} U^{(k)}_{\Lambda} &=& \Big[ \frac{\partial^{k}}{\partial \rho^k} \big(\partial_{\Lambda} U_{\Lambda}(\rho)\big) -\frac{U^{(k+1)}_{\Lambda}}{U_{\Lambda}^{(2)}} \frac{\partial}{\partial \rho} \big(\partial_{\Lambda} U_{\Lambda}(\rho)\big) \Big]
_{\rho^0_{\Lambda}},
 \nonumber
 \\
 & &
\end{eqnarray}
The flow of the couplings $Z_{\Lambda}, Y_{\Lambda}$ and $V_{\Lambda}$ related
to the single-particle Green function can be derived
by inserting our ansatz (\ref{eq:Gammagrad}) into the exact FRG flow 
equation for $\Gamma_{\Lambda} [ \bar{\phi} , \phi ]$ and comparing terms 
with the same number of gradients on both sides.~\cite{Wetterich93,Berges02} 
If we retain only the  quadratic coupling $U^{(2)}_{\Lambda} \equiv u_{\Lambda} $
and set $ Y_{\Lambda} =1$ and $V_{\Lambda}=0$ we recover the flow equations
(\ref{eq:rhoflowT}--\ref{eq:etadef})
obtained in Sec.~\ref{sec:FRGvert} within the vertex expansion.
If we retain in addition the cubic and quartic couplings
 $U^{(3)}_{\Lambda}$ and $U^{(4)}_{\Lambda}$ the resulting system of equations 
is lather lengthy. 
In this work we do not give these equations explicitly because
their derivation is  straightforward  and 
they can only be analysed numerically anyway; technical details can be found in
Ref.~[\onlinecite{Eichler09}].
Note that the
couplings $U_{\Lambda}^{(k)}$ and $V_{\Lambda}$ are not dimensionless;
for our numerical analysis it is convenient to work with the corresponding
dimensionless couplings 
 \begin{equation}
 \tilde{U}_{\Lambda}^{(k)} = 2 m U_{\Lambda}^{(k)}  
\Lambda_0^{D(k-1) -2}   \; \; \; , \; \; \; 
\tilde{V}_{\Lambda} = \frac{ \Lambda_0^2 V_{\Lambda}}{2m} .
 \label{eq:tildeU234def} 
\end{equation}
For simplicity, we have explicitly solved the
coupled flow equations for the seven dimensionless couplings
$\tilde{U}_{\Lambda}^{(2)}, 
\tilde{U}_{\Lambda}^{(3)}, 
\tilde{U}_{\Lambda}^{(4)},
\tilde{\rho}^{0}_{\Lambda},
Z_{\Lambda},
Y_{\Lambda}$ and $\tilde{V}_{\Lambda}$
only at zero temperature. The typical RG flow for the
coupling constants $\tilde{U}_{\Lambda}^{(2)}, 
\tilde{U}_{\Lambda}^{(3)}$ and
$\tilde{U}_{\Lambda}^{(4)}$ in two dimensions is shown in Fig.~\ref{fig:U234flow}.
Obviously, during the RG flow the cubic and quartic couplings
$\tilde{U}_{\Lambda}^{(3)}$ and
$\tilde{U}_{\Lambda}^{(4)}$ become orders of magnitude larger
than the quadratic coupling $\tilde{U}_{\Lambda}^{(2)}$.
Keeping in mind that a  low order truncated polynomial
approximation for the effective potential is only justified
if the neglected higher order terms 
are smaller than the retained lower order terms, we conclude from
Fig.~\ref{fig:U234flow} that  results based on truncations of the derivative expansion with
quadratic effective potential should be considered with some skepticism.
This is also the case in three dimensions where a similar calculation (not shown here)
leads to a divergence of both coefficients  $\tilde{U}_{\Lambda}^{(3)}$ and $\tilde{U}_{\Lambda}^{(4)}$ for $\Lambda \rightarrow 0$.

\begin{figure}[h]
\centering
\includegraphics[width=8cm]{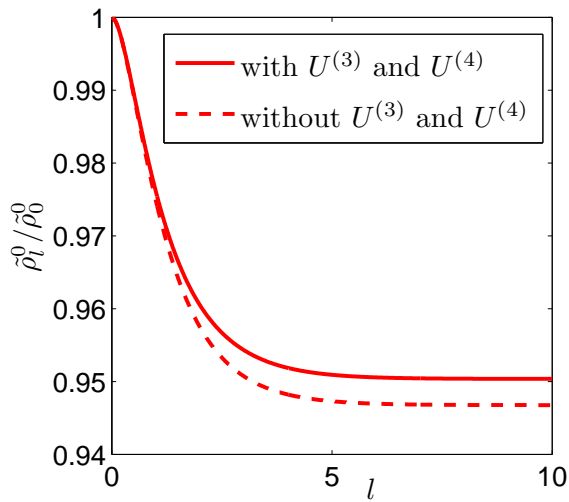}
\caption{(Color online)  
RG flow of the condensate density for $\tilde{u}_0=3$, $\tilde{\mu}=1$, $D=2$ and $T=0$ with and without the
cubic and quartic couplings 
$\tilde{U}_{\Lambda}^{(3)}$ and
$\tilde{U}_{\Lambda}^{(4)}$.
} 
\label{fig:condens}
\end{figure}

In spite of this, the widely
used quadratic approximation for the effective potential
can still give acceptable results for some physical quantities.
For example, the large values of
$\tilde{U}_{\Lambda}^{(3)}$ and
$\tilde{U}_{\Lambda}^{(4)}$ shown in  Fig.~\ref{fig:U234flow}
lead only to a small correction for the condensate density, as shown in
Fig.~\ref{fig:condens}.
Interestingly, the inclusion of cubic and quartic terms
in the expansion of the effective potential
renormalize the condensate density to larger values, so that
the quadratic approximation overestimates
the effect of fluctuations.
It would be interesting to determine the true form 
of the effective potential in two and three dimensions
by directly solving the partial differential equation (\ref{eq:Udif}).
This seems to be a rather difficult numerical task which is beyond the scope of this work.

\section{Summary and outlook}

In summary, we have used two 
different truncation strategies for the formally exact FRG flow equation
for the generating functional of the irreducible vertices to
calculate the condensate density of the interacting Bose gas.
Our first strategy presented in Sec.~\ref{sec:FRGvert} 
is based on the truncated vertex expansion
recently proposed in Ref.~[\onlinecite{Sinner09}].
We have further simplified  this truncation at finite temperature 
to obtain a closed system of
flow equations for the condensate density, the effective interaction, and
the wave-function renormalization factor at finite temperature.
These flow equations are equivalent to the flow equations recently derived by
Floerchinger and Wetterich within the derivative expansion \cite{Floerchinger09}.
From the numerical solution of these flow equations
we have obtained a quantitatively accurate
description of the critical regime.
We have also developed 
an extrapolation procedure to
extract the order parameter exponent $\beta$ with high accuracy
and moderate computational effort from the numerical solution of the
FRG flow equations.

In order to estimate
the renormalization of the condensate density by
three-body and four-body interactions which are not included
in our truncated vertex expansion of Sec.~\ref{sec:FRGvert},
we have used in Sec.~\ref{sec:derivative}
the truncated derivative expansion in combination
with a polynomial approximation for the effective potential
and derived appropriate FRG flow equations at zero temperature.
We have shown that 
the RG flow drives the cubic and quartic coefficients
 $\tilde{U}_{\Lambda}^{(3)}$ and
$\tilde{U}_{\Lambda}^{(4)}$
in the expansion 
of the effective potential $U_{\Lambda}(\rho)$
to rather large values.
This indicates that a low order polynomial approximation
is not appropriate to describe the
true form of the effective potential in the condensed phase.
Although the inclusion of the  
coefficients $\tilde{U}_{\Lambda}^{(3)}$ and
$\tilde{U}_{\Lambda}^{(4)}$ leads only to a small positive correction of the 
condensate density, 
other physical observables might be more strongly affected by
the three- and four-body interactions described by these couplings.

By using both the vertex expansion and the derivative expansion, we have
clearly established the relation between these approximation strategies and have
illustrated their  advantages.
In this work we have focused on the 
condensate density, because
it appears naturally as one of the flowing couplings 
in the FRG flow equations.
Our work can be extended in several directions: First of all,
by keeping track of the
field-independent part $\Gamma_{\Lambda}^{(0)}$
of the generating functional in Eq.~(\ref{eq:defgamma})
(which can be identified with the interaction correction
to the grand canonical potential per unit volume)
one can also obtain the FRG flow of 
any thermodynamic observable of interest\cite{Floerchinger09}. 
In the critical regime  our numerical extrapolation procedure
outlined in Sec.~\ref{sec:FRGvert} should
again be useful to obtain quantitatively accurate results
for other critical exponents.
It would also be interesting to generalize 
the vertex expansion approach developed in Ref.~[\onlinecite{Sinner09}]
to finite temperatures and calculate the
single-particle spectral function in the critical regime.

\section*{ACKNOWLEDGMENTS}
We thank Andreas Sinner for discussions and
gratefully acknowledge financial support by
SFB/TRR49 and the DAAD/CAPES PROBRAL-program.


\begin{thebibliography}{99}
\bibitem{footnotegenerating}
We normalize the generating functional 
$\Gamma_{\Lambda} [ \bar \phi , \phi  ]$
of the irreducible vertices in Eq.~(\ref{eq:gendef})
such that at the initial RG scale
it reduces to the bare interaction.
The second derivative of
$\Gamma_{\Lambda} [ \bar \phi , \phi  ]$ can then be identified with the
irreducible self-energy.
In contrast, the corresponding generating functional
used by Wetterich and co-authors 
\cite{Wetterich08,Floerchinger09,Wetterich93,Berges02}
is normalized such that its second derivative is the exact inverse propagator.
For a detailed discussion of the advantages of different normalizations see
Ref.~[\onlinecite{Kopietz09}].
\bibitem{footnotemistake}
There is a printing error in Eq.~(7) of Ref.~[\onlinecite{Sinner09}]:
the density fluctuations $\delta \rho_K$ should be replaced by $\rho_K$.
\end{thebibliography}
\end{document}